\documentclass[twocolumn,showpacs,aps,prl,superscriptaddress]{revtex4}

\usepackage{graphicx}
\usepackage{dcolumn}
\usepackage{amsmath}
\usepackage{epsfig}

\input babarsym.tex
\newcommand{\bi}{\begin{itemize}}
\newcommand{\ei}{\end{itemize}}
\newcommand{\ben}{\begin{enumerate}}
\newcommand{\een}{\end{enumerate}}
\newcommand{\bc}{\begin{center}}
\newcommand{\ec}{\end{center}}
\newcommand{\bt}{\begin{table}}
\newcommand{\et}{\end{table}}
\newcommand{\be}{\begin{equation}}
\newcommand{\eeq}{\end{equation}}
\newcommand{\ba}{\begin{eqnarray}}
\newcommand{\ea}{\end{eqnarray}}

\newcommand{\la}{\ifmmode {\leftarrow} \else {$\leftarrow$}\fi}
\newcommand{\Ra}{\ifmmode {\Rightarrow} \else {$\Rightarrow$}\fi}
\newcommand{\La}{\ifmmode {\Leftarrow} \else {$\Leftarrow$}\fi}
\newcommand{\Lra}{\ifmmode {\Longrightarrow} \else {$\Longrightarrow$}\fi}
\newcommand{\Lla}{\ifmmode {\Longleftarrow} \else {$\Longleftarrow$\fi}}
\newcommand{\Llra}{\ifmmode {\Longleftrightarrow} \else {$\Longleftrightarrow$\fi}}
\newcommand{\Lk}{\ifmmode {{\cal L}} \else {${\cal L}$}\fi}
\newcommand{\Wt}{\ifmmode {{\cal W}} \else {${\cal W}$}\fi}
\newcommand{\Br}{\ifmmode {{\cal B}} \else {${\cal B}$}\fi}
\newcommand{\N}{\ifmmode {{\cal N}} \else {${\cal N}$}\fi}
\newcommand{\G}{\ifmmode {{\cal G}} \else {${\cal G}$}\fi}
\newcommand{\E}{\ifmmode {{\cal E}} \else {${\cal E}$}\fi}

\newcommand{\tBz}{\ifmmode {\tau_{\Bz}} \else {$\tau_{\Bz}$ }\fi }
\newcommand{\tBp}{\ifmmode {\tau_{\Bu}} \else {$\tau_{\Bu}$ }\fi }
\def\nubar    {\kern 0.18em\overline{\kern -0.18em \nu}{}\xspace}
\def\nulb     {\ensuremath{\nubar_\ell}\xspace}

\newcommand{\psoft}{\ifmmode {{\pi_s}^+} \else {${\pi_s}^+$}\fi }
\newcommand{\dm}{\ifmmode {\Delta M} \else {$\Delta M$}\fi}
\newcommand{\plab}{\ifmmode{p} \else {$p$}\fi}
\newcommand{\ks}{\ifmmode{k^*} \else {$k^*$}\fi}
\newcommand{\om}{\ifmmode{w} \else {$w$}\fi}
\newcommand{\omt} {\ifmmode {\tilde{w}} \else {$\tilde{w}$} \fi}
\newcommand{\mnusq}{\ifmmode{{M_\nu}^2} \else {${M_{\nu}}^2$}\fi} 
\newcommand{\DTau}{\ifmmode {\Delta \tau} \else {$\Delta \tau$}\fi}
\newcommand{\ggcc}{\ifmmode {GeV^2/c^4} \else {$GeV^2/c^4$}\fi}
\newcommand{\TBY}{\ifmmode{\theta_{\Bz, D^*\ell}} \else {$\theta_{\Bz, D^*\ell}$} \fi}
\newcommand{\Aone}{\ifmmode {{\cal A}_1} \else {${\cal A}_1$}\fi}
\newcommand{\rha}{\ifmmode{\mbox{\rho^2_{{\cal A}_1}}} \else {\mbox{$\rho^2_{{\cal A}_1}$}}\fi}
\def\BpBm {\ensuremath{B^+ {\kern -0.16em \Bub}}}

\newcommand{\BABARPubYear}    {04}
\newcommand{\BABARPubNumber}  {007}

\newcommand{\SLACPubNumber} {10403}

\def\figurebox#1#2#3{%
    \def\arg{#3}%
    \ifx\arg\empty
    {\hfill\vbox{\hsize#2\hrule\hbox to #2{\vrule\hfill\vbox to #1{\hsize#2\vfill}\vrule}\hrule}\hfill}%
    \else
    {\hfill\epsfbox{#3}\hfill}%
    \fi}

\begin{document}

\begin{flushleft}
\babar-PUB-\BABARPubYear/\BABARPubNumber\\
SLAC-PUB-\SLACPubNumber\\
\end{flushleft}

\title{

{\large {
Determination of the Branching Fraction for ${\boldmath B \rightarrow X_c \ell \nu}$ Decays and of ${\boldmath |V_{cb}|}$ from Hadronic-Mass and Lepton-Energy Moments }}}

%
\author{B.~Aubert}
\author{R.~Barate}
\author{D.~Boutigny}
\author{F.~Couderc}
\author{J.-M.~Gaillard}
\author{A.~Hicheur}
\author{Y.~Karyotakis}
\author{J.~P.~Lees}
\author{V.~Tisserand}
\author{A.~Zghiche}
\affiliation{Laboratoire de Physique des Particules, F-74941 Annecy-le-Vieux, France }
% ...
\collaboration{The \babar\ Collaboration}
\noaffiliation

\date{\today}

\begin{abstract}

We determine the inclusive $B \rightarrow X_c \ell \nu$ branching fraction, ${\cal B}_{c\ell\nu}$,
the CKM matrix element $|V_{cb}|$, and other heavy-quark parameters from a simultaneous 
fit to moments of the hadronic-mass and lepton-energy distributions in semileptonic $B$-meson decays, measured as a function of the lower limit on the lepton energy, using data recorded with the \babar\ detector. Using Heavy Quark Expansions (HQEs) 
to order $1/m_b^3$, we extract ${\cal B}_{c e \nu}=(10.61 \pm 0.16_{exp} \pm 0.06_{HQE}) \% $ and   $|V_{cb}|= (41.4 \pm 0.4_{exp} \pm  0.4_{HQE} \pm 0.6_{th})\times 10^{-3}$. The stated errors refer to the experimental, HQE, and additional theoretical uncertainties.
\end{abstract}

\maketitle

The CKM matrix element $V_{cb}$ is one of the fundamental parameters of the Standard Model and
thus its precise measurement with well understood uncertainties is important. 
At the parton level, the weak decay rate for $b\ra c \ell \nu$  can be calculated 
accurately; it is proportional to $|V_{cb}|^2$ and depends on the quark masses,
$m_b$ and $m_c$.  To relate 
measurements of the semileptonic $B$-meson decay rate 
to $|V_{cb}|$, the parton-level calculations 
have to be corrected for effects of strong interactions.
Heavy-Quark Expansions (HQEs)\cite{hqe1} have become a useful tool for calculating 
perturbative and non-perturbative QCD 
corrections~\cite{hqe2} and for estimating 
their uncertainties.  
In the kinetic-mass scheme~\cite{kinetic} for example, these expansions in $1/m_b$ 
and $\alpha_s(m_b)$ (the strong coupling constant) 
to order ${\cal O}(1/m_b^3)$ contain six parameters: the running kinetic masses of the $b-$ and 
$c-$quarks, $m_b(\mu)$ and $m_c(\mu)$, and four non-perturbative parameters.
We determine these parameters 
from a fit to the moments of the hadronic-mass and electron-energy distributions in semileptonic $B$ decays to charm particles, $B \ra X_c \ell \nu$. 
This fit yields measurements of the inclusive branching fraction ${\cal B}_{c \ell \nu}$
and of $|V_{cb}|$, significantly improved compared to earlier \babar\ measurements~\cite{thorsten}.
It also allows us to test the consistency of the data with the HQEs employed and
to check for the possible impact of higher-order contributions. 
Moment measurements and the extraction of $|V_{cb}|$ based on HQEs~\cite{vol-falk} were first performed by the CLEO collaboration~\cite{cleo}. More recently, global fits to a variety of moments were presented~\cite{bauer,cleo2,delphi}, using HQEs in different mass schemes.

\begin{figure*}[t]
\begin{centering}
\epsfig{file=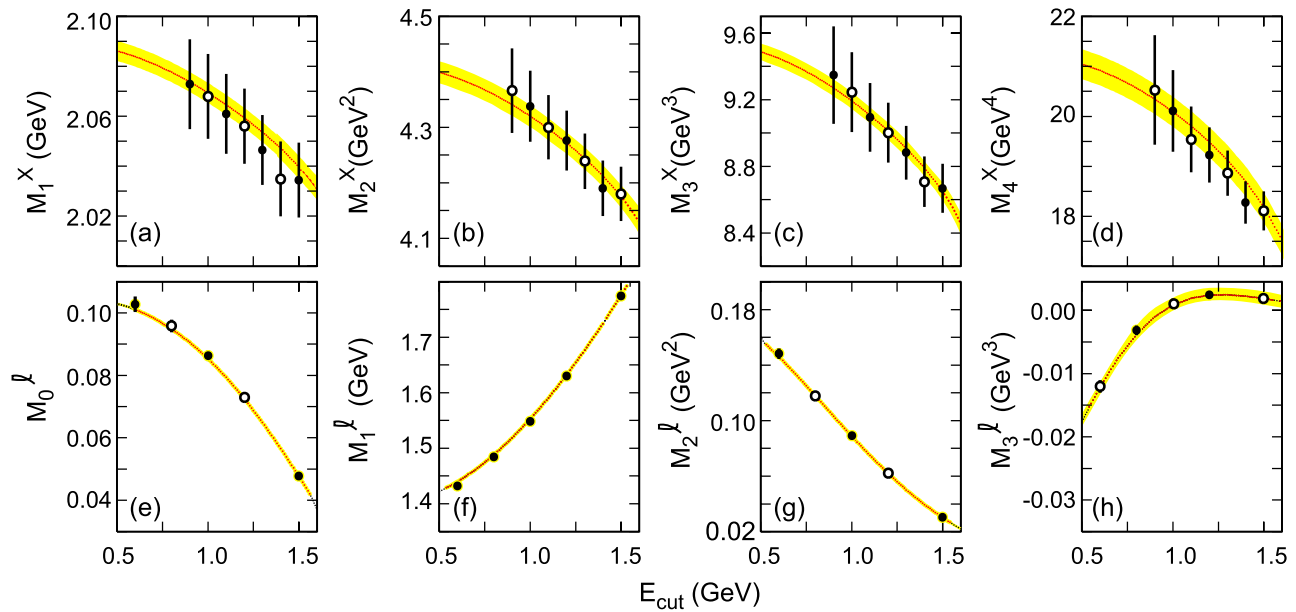, width=17cm}
\caption{The measured hadronic-mass (a-d) and electron-energy (e-h) moments as a function of the cut-off energy, $E_{cut}$,
compared with the result of the simultaneous fit (line), with the theoretical uncertainties~\cite{gambino} indicated as shaded bands.  The solid data points mark the measurements included in the fit.  The vertical bars indicate the experimental errors;
in some cases they are comparable in size to the data points.  Moment measurements for different $E_{cut}$ are highly correlated.
\label{fig:moments}}
\end{centering}
 \end{figure*}

This analysis makes use of moments measured by the \babar\ collaboration~\cite{emoments,mxmoments}.  The moments are
derived from the inclusive hadronic-mass ($m_X$) and electron-energy ($E_{\ell}$) distributions in $B \ra X_c \ell \nu$ decays, averaged over charged and neutral $B$ mesons produced at the $\FourS$ resonance. We have subtracted the charmless contributions 
based on the branching fraction ${\cal B}_{u \ell \nu}$=$(0.22 \pm 0.05)\%$~\cite{btou}.
All moments are measured as functions of $E_{cut}$, a lower limit on the lepton energy (for energy moments we only use electrons, for mass moments we also use muons).  The moments are corrected for detector effects and QED radiation~\cite{photos}.
The hadronic-mass distribution is measured in events tagged by the fully reconstructed hadronic decay of the second $B$ meson.  The hadronic-mass moments are defined as
$M_n^X(E_{cut})=\langle m_X^n \rangle_{E_{\ell}>E_{cut}}$ with $\emph{n}=$1,2,3,4.
The electron-energy distribution is measured in events tagged by a high-momentum electron from the second $B$ meson.  We define 
the first energy moment as $M_1^{\ell}(E_{cut}) = \langle E_{\ell}\rangle_{E_{\ell}>E_{cut}}$  and 
the second and third moments as
$M_n^{\ell}(E_{cut})=\langle(E_{\ell} - M_1^{\ell}(E_{cut}))^n\rangle_{E_{\ell}>E_{cut}}$ 
with $\emph{n}=$2,3. 
In addition, we use the partial branching fraction 
$M_0^{\ell}(E_{cut})=\int_{E_{cut}}^{E_{max}}(d{\cal B}_{c \ell \nu}/dE_{\ell})\, dE_{\ell} $.
All measured moments are shown in Fig.~\ref{fig:moments}. 

In the kinetic-mass scheme the HQE to ${\cal O}(1/m_b^3)$ for the rate of $B \ra X_c \ell \nu$ decays can be expressed as~\cite{kolya} 
\begin{eqnarray}
\label{equ:opegsl}
\nonumber
\Gamma_{c \ell \nu} = \frac{G_F^2 m_b^5}{192\pi^3} |V_{cb}|^2 (1+A_{ew}) A_{pert}(r,\mu) \times \nonumber \\
\Bigg [ z_0(r) \Bigg ( 1 - \frac{\mu_{\pi}^2-\mu_G^2+\frac{\rho_D^3+\rho_{LS}^3}{m_b}}{2m_b^2} \Bigg ) \nonumber \\
 - 2(1-r)^4\frac{\mu_G^2+\frac{\rho_D^3+\rho_{LS}^3}{m_b}}{m_b^2}+d(r)\frac{\rho_D^3}{m_b^3}+{\cal O}(1/m_b^4)\Bigg].
\end{eqnarray}
\noindent
The leading non-perturbative effects
arise at ${\cal O}(1/m_b^2)$ and are parameterized by $\mu_{\pi}^2(\mu)$ and $\mu_{G}^2(\mu)$,  the expectation values of the kinetic and chromomagnetic dimension-five operators. 
At ${\cal O}(1/m_b^3)$, two additional parameters enter, 
$\rho_{D}^3(\mu)$ and $\rho_{LS}^3(\mu)$, the  expectation values 
of the Darwin ($D$) and spin-orbit ($LS$) dimension-six operators. 
These parameters depend on the scale $\mu$ that  separates short-distance from long-distance QCD effects;  
the calculations are performed for $\mu=1\gev$~\cite{kinetic}.
Electroweak corrections are 
$1+A_{ew} \cong ( 1 + \alpha/\pi \, \mathrm{ln}(M_Z/m_b) )^2 = 1.014 $ 
and  perturbative QCD corrections are estimated to be
$A_{pert}(r,\mu)\cong 0.91\pm 0.01$~\cite{kolya}. 
The ratio $r=m_c^2/m_b^2$ enters in the phase-space factor  $z_0(r)=1-8r+8r^3-r^4-12r^2\,\mathrm{ln}r$
and the function $d(r)= 8\,\mathrm{ln}r + 34/3 - 32r/3 - 8r^2 + 32r^3/3 - 10r^4 /3  $.

This analysis uses linearized expressions for the HQEs~\cite{gambino}. Specifically, the dependence of $|V_{cb}|$ 
on the true values of heavy-quark parameters, expanded around ${\it a~priori}$ estimates of these parameters, is
\begin{eqnarray}
\nonumber
& &\frac{|V_{cb}|}{0.0417} \cong \sqrt{\frac{{\cal B}_{clv}}{(0.1032)} 
\frac{1.55\,\mathrm{ps}}{\tau_{B}} }  \\
& &\times [1+0.30 (\alpha_s(m_b)-0.22) ] \nonumber \\ 
& &\times [ 1 - 0.66 ( m_b -4.60) +0.39 ( m_c -1.15 ) \nonumber \\
& &+ 0.013 ( \mu_{\pi}^2 -0.40) +0.09 ( \rho_D^3 - 0.20) \nonumber \\
& &+ 0.05 ( \mu_G^2-0.35 ) -0.01 ( \rho_{LS}^3+0.15 ) ]. 
\label{equ:opevcb}
\end{eqnarray}
Here $m_b$ and all other parameters of the expansion are in $\gev^{(n)}$;
$\tau_B$ refers to the average lifetime of $B$ mesons produced at the $\FourS$.
We use $\tau_B = f_0 \tau_0 +(1-f_0) \tau_{\pm} = (1.608 \pm 0.012)\,$ps, taking into account the lifetimes~\cite{pdg2002} of neutral and charged $B$ mesons, $\tau_0$ and $\tau_{\pm}$, and their relative production rates (defined in terms of $f_0=0.488\pm 0.013$~\cite{f0}, the fraction of $\Bz\Bzb$ pairs). 

HQEs in terms of the same heavy-quark parameters are available for the hadronic-mass and electron-energy moments.
The dependence on the heavy-quark parameters has again been linearized using the same ${\it a~priori}$  estimates of the parameters~\cite{kolya,gambino}. 
We have verified that the differences between the linearized expressions and the full theoretical calculation are very small in all cases.
We use these linear equations to determine the unknown heavy-quark parameters, the total branching fraction ${\cal B}_{c \ell \nu}$, and $|V_{cb}|$ from a simultaneous $\chi^2$ fit to the measured moments and the partial branching fraction, all as a function of the cut-off lepton energy,  $E_{cut}$. 

In total, we have measured four hadronic-mass moments for each of seven different values of $E_{cut}$, ranging from 0.9 to 1.5 \gev, and three electron-energy moments plus the partial branching fraction at five values of $E_{cut}$, ranging from 0.6 to 1.5\gev~\cite{emoments,mxmoments}.  Since many of these individual moments are highly correlated we select for the fitting procedure a set of moments for which the correlation coefficients do not exceed 95\%.  Thus we only use half of the 28 mass moments, 
and retain 13 of the 20 energy moments.

The global fit takes into account the statistical and systematic errors and correlations of the individual measurements, as well as the 
uncertainties of the expressions for the individual moments.
We assess the uncertainty of the calculated moments by varying, as suggested in~\cite{gambino}, in the linearized expressions (given for $|V_{cb}|$ in Eq.~\ref{equ:opevcb}) the ${\it a~priori}$  estimates for
$\mu_{\pi}^2$ and $\mu_G^2$  by $\pm 20\%$ and for $\rho_{LS}^3$ and $\rho_D^3$ by $\pm 30\%$. 
We assume that for a given moment these changes are fully correlated for all values of $E_{cut}$, but uncorrelated for different moments.
The resulting fit, shown in Fig.~\ref{fig:moments}, describes the data well with $\chi^2=15.0$ for 20 degrees of freedom.  
Tables~\ref{tab:results} and  \ref{tab:correlations} list the fitted parameters, their errors and correlations. As expected, $m_b$ and $m_c$ are highly correlated.
For the mass difference, we obtain $ m_b - m_c=(3.436 \pm 0.025_{exp} \pm 0.018_{HQE} \pm 0.010_{\alpha_s}) \gev$. 

Beyond the uncertainties 
that are included in the fit, 
the limited knowledge of the expression for the decay rate, including various perturbative corrections and higher-order non-perturbative corrections, introduces an error in $|V_{cb}|$, assessed to be 1.5\%~\cite{kolya}.
By comparison, the impact of the uncertainty in  $\alpha_s$  is estimated to be relatively small.  
For $M^{\ell}_n(E_{cut})$ moments, perturbative corrections of order $\alpha_s^2$ are included with $\alpha_s(m_b)=0.22\pm 0.04$,  
whereas for $M^X_n(E_{cut})$ moments, they are calculated only to ${\cal O}(\alpha_s)$
with $\alpha_s(m_b)=0.3\pm 0.1$.
We estimate the error on the perturbative corrections by varying $\alpha_s$ within the stated errors.  The choice of the scale $\mu$ is estimated to have a very small impact on $|V_{cb}|$ and the branching fraction~\cite{kolya}.
\begin{table*}[htbp]
\caption[]
{
\label{tab:results} Fit results and error contributions from the 
moment measurements, approximations to the HQEs, and additional theoretical uncertainties from $\alpha_s$ terms and other perturbative and non-perturbative terms contributing to $\Gamma_{c\ell\nu}$.
}
\renewcommand{\arraystretch}{1.10}
\begin{tabular}{lD{.}{.}{-1}D{.}{.}{-1}D{.}{.}{-1}D{.}{.}{-1}D{.}{.}{-1}D{.}{.}{-1}D{.}{.}{-1}D{.}{.}{-1}}
\hline\hline
&\multicolumn{1}{r}{$|V_{cb}| (10^{-3})$} 
&\multicolumn{1}{r}{$m_b $ $\mathrm{(GeV)}$} 
&\multicolumn{1}{r}{$m_c $ $\mathrm{(GeV)}$} 
&\multicolumn{1}{r}{$\mu_{\pi}^2 $ $\mathrm{(GeV^2)}$} 
&\multicolumn{1}{r}{$\rho_D^3 $ $\mathrm{(GeV^3)}$} 
&\multicolumn{1}{r}{$\mu_{G}^2 $ $\mathrm{(GeV^2)}$} 
&\multicolumn{1}{r}{$\rho_{LS}^3 $ $\mathrm{(GeV^3)}$} 
&\multicolumn{1}{r}{${\cal B}_{c\ell\nu} $ $(\%)$}    \\ \hline
             Results &41.390 &4.611 &1.175  &0.447 &0.195 &0.267 &-0.085 &10.611    \\ \hline 
        $\delta_{exp}$ & 0.437 &0.052 &0.072 &0.035 &0.023 &0.055 &0.038 & 0.163 \\ 
        $\delta_{HQE}$ & 0.398 &0.041 &0.056 &0.038 &0.018 &0.033 &0.072 & 0.063 \\ 
   $\delta_{\alpha_s}$ & 0.150 &0.015 &0.015 &0.010 &0.004 &0.018 &0.010 & 0.000 \\ 
   $\delta_{\Gamma}  $ & 0.620 &      &      &      &      &      &      &       \\ \hline
        $\delta_{tot}$ & 0.870 &0.068 &0.092 &0.053 &0.029 &0.067 &0.082 & 0.175 \\ \hline\hline
\end{tabular}
\end{table*}

\begin{table}[htb]
\caption[]
{
\label{tab:correlations} Correlation coefficients for the fit parameters.
}
\renewcommand{\arraystretch}{1.10}
\begin{tabular}{lD{.}{.}{-1}D{.}{.}{-1}D{.}{.}{-1}D{.}{.}{-1}D{.}{.}{-1}D{.}{.}{-1}D{.}{.}{-1}D{.}{.}{-1}}
\hline\hline
&\multicolumn{1}{r}{$|V_{cb}|$}  
&\multicolumn{1}{r}{$m_b$} 
&\multicolumn{1}{r}{$m_c$} 
&\multicolumn{1}{r}{$\mu_{\pi}^2$} 
&\multicolumn{1}{r}{$\rho_D^3$}  
&\multicolumn{1}{r}{$\mu_{G}^2$}  
&\multicolumn{1}{r}{$\rho_{LS}^3$}  
&\multicolumn{1}{r}{${\cal B}_{c\ell\nu}$}   \\ \hline
$|V_{cb}|$             & 1.00&-0.49&-0.36& 0.56& 0.35&-0.37& 0.64& 0.61 \\
$m_b$                  &     & 1.00& 0.97&-0.40&-0.13& 0.16&-0.63& 0.23 \\
$m_c$                  &     &     & 1.00&-0.38&-0.13&-0.04&-0.50& 0.29\\
$\mu_{\pi}^2 $         &     &     &     & 1.00& 0.82& 0.08& 0.46& 0.16\\
$\rho_D^3  $           &     &     &     &     & 1.00& 0.08& 0.23& 0.12\\
$\mu_{G}^2  $          &     &     &     &     &     & 1.00&-0.43&-0.04\\
$\rho_{LS}^3 $         &     &     &     &     &     &     & 1.00& 0.09\\ 
${\cal B}_{c\ell\nu}$  &     &     &     &     &     &     &     & 1.00\\
\hline\hline
\end{tabular}
\end{table}

A series of tests has been performed to verify that the fit results are unbiased.
Specifically, we enlarged and reduced the estimated theoretical uncertainties by a factor of two
and verified that the changes in the fitted parameters were small compared to the errors of the standard fit. 
We have also checked that the choice of the set of moments that are used in the fit does not significantly affect the result.  In particular, an energy cut-off
above 1.2 \gev\ might have introduced larger theoretical uncertainties and a potential bias, and moments for lower values of $E_{cut}$ might have been affected by higher backgrounds.  We found no evidence for any such effects.

The fit results are fully compatible with 
independent estimates~\cite{gambino} of $\mu_G^2$=$(0.35 \pm 0.07)\gev^2$, based on the 
$B^* - B$ mass splitting, and of $\rho_{LS}^3$=$(-0.15 \pm 0.10)\gev^3$, from heavy-quark sum 
rules~\cite{sumrules}.

\begin{figure}[!t]
\begin{center}
\epsfig{file=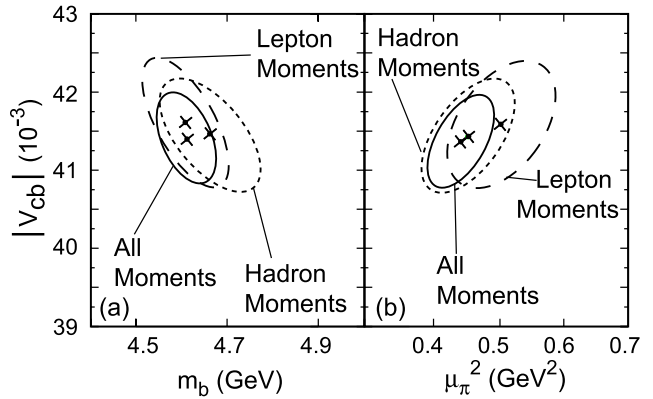, width=8.5cm}
\caption{
Fit results (crosses) with contours corresponding to $\Delta \chi^2=1$ for two pairs of the eight free parameters
a) $m_b$ and b) $\mu_{\pi}^2$ versus $|V_{cb}|$, separately for fits using the hadronic-mass, the electron-energy, and all moments.
}
\label{fig:ellipses}
\end{center}
\end{figure}

Figure~\ref{fig:ellipses} shows the $\Delta \chi^2=1$ ellipses for $|V_{cb}|$ versus $m_b$ and $\mu_{\pi}^2$, for a fit to all moments and separate fits to the electron-energy moments and the hadronic-mass moments, but including the partial branching fractions in both. 
The lepton-energy and hadronic-mass moments have slightly different sensitivity to the fit parameters, but the results for the separate fits, $|V_{cb}|=(41.4 \pm 0.7)\,\times 10^{-3}$ and $|V_{cb}|=(41.6 \pm 0.8)\,\times 10^{-3}$ are fully compatible with each other and with the global fit to all moments. Changes in the other fit parameters are also consistent within the stated errors.
Since the expansions for the two sets of moments are sensitive to different theoretical uncertainties and assumptions, in particular the differences in the treatment of the perturbative corrections, the observed consistency of the separate fits indicates that such differences are small compared with the experimental and assumed theoretical uncertainties.

In conclusion, we have extracted $|V_{cb}|$, the semileptonic branching fraction, and the 
heavy-quark masses,
\begin{eqnarray}
\nonumber
|V_{cb}|&=& (41.4 \pm 0.4_{exp} \pm  0.4_{HQE} \pm 0.6_{th})\,\times 10^{-3},  \nonumber \\
{\cal B}_{c e \nu} &=& ( 10.61 \pm 0.16_{exp} \pm 0.06_{HQE}) \%, \nonumber  \\
m_b(1 \gev)&=&(4.61 \pm 0.05_{exp} \pm 0.04_{HQE} \pm 0.02_{th}) \gev, \nonumber \\
m_c(1 \gev)&=&(1.18 \pm 0.07_{exp} \pm 0.06_{HQE} \pm 0.02_{th}) \gev, \nonumber
\end{eqnarray}
as well as the non-perturbative parameters in the kinetic-mass scheme up to order $(1/m_b^3)$
(see Table~\ref{tab:results}).  
The total semileptonic branching fraction is ${\cal B}_{c e \nu} + {\cal B}_{u e \nu} = ( 10.83 \pm 0.16_{exp} \pm 0.06_{HQE})\%$.
The errors refer to contributions from the experimental errors on the moment measurements, the HQE uncertainties included in the fit, and additional theoretical uncertainties,
$\delta_{th}=\sqrt{\delta_{\alpha_s}^2 + \delta_{\Gamma}^2}$, derived from Refs.~\cite{kolya,gambino}.

Based on a large set of hadronic-mass and electron-energy moments and  a consistent set of HQE calculations, we have also been able to assess the uncertainties in the ${\cal O}(1/m_b^3)$ terms from the data without constraints to any {\it a priori} values. The fitted values of the parameters are consistent with theoretical estimates~\cite{kinetic,kolya}. The uncertainties on the quark masses are much smaller than those of previous measurements~\cite{pdg2002}.
It would be interesting to
compare the results of this analysis with fits based on recent calculations performed in the $1S$ mass scheme~\cite{trott}. 

The result on $|V_{cb}|$ is in agreement with previous measurements using HQEs, either for a different mass scheme and with fixed terms of ${\cal O}(1/m_b^3)$~\cite{cleo2},
or for the kinetic-mass scheme, but with external constraints on almost all HQE parameters~\cite{delphi}, as well as with an analysis combining both of these measurements~\cite{bauer}.

This work has greatly benefited from many interactions with  N. Uraltsev and P. Gambino. We thank them as well as many other theorists for valuable discussions regarding this work.
We are grateful for the excellent luminosity and machine conditions
provided by our \pep2\ colleagues, 
and for the substantial dedicated effort from
the computing organizations that support \babar.
The collaborating institutions wish to thank 
SLAC for its support and kind hospitality. 
This work is supported by
DOE
and NSF (USA),
NSERC (Canada),
IHEP (China),
CEA and
CNRS-IN2P3
(France),
BMBF and DFG
(Germany),
INFN (Italy),
FOM (The Netherlands),
NFR (Norway),
MIST (Russia), and
PPARC (United Kingdom). 
Individuals have received support from the 
A.~P.~Sloan Foundation, 
Research Corporation,
and Alexander von Humboldt Foundation.

\end{document}